\documentstyle[sprocl]{article}

\def\Journal#1#2#3#4{{#1} {\bf #2}, #3 (#4)}
\def\NPB{{\em Nucl. Phys.} B}

\def\PRL{\em Phys. Rev. Lett.}
\def\PRD{{\em Phys. Rev.} D}

\def\be{\begin{equation}}
\def\ee{\end{equation}}
\def\bea{\begin{eqnarray}}
\def\eea{\end{eqnarray}}

\begin{document}

\title{Generalized Parton Distributions and Distribution of Partons in
the Transverse Plane
}

\author{M. Burkardt}

\address{Department of Physics, New Mexico State University,
Las Cruces, NM 88011, USA\\
E-mail: burkardt@nmsu.edu}

%\author{Elsie Tan, Jessie Tan and R. Sankaran}%
%
%\address{World Scientific Publishing Co Ltd,
%57 Shelton Street, Covent Garden, London WC2H 9HE, England\\
%E-mail: wspc@wspc.ox.uk}

\maketitle

\abstracts{
I discuss the physical interpretation of GPDs in the limit $\xi=0$,
where the $t$ dependence contains information about the
distribution of partons in the transverse plane. GPDs thus allow
a simultaneous determination of the longitudinal momentum and
transverse position of partons in the infinite momentum frame,
which also provides a physical interpretation for Ji's angular 
momentum sum rule.
}

\section{Motivation}
Recently, there has been a lot of interest in
Generalized parton distributions (GPDs) because
on the one hand GPDs can be probed in real and
deeply virtual Compton scattering (DVCS) and on the other hand they
can be related to the total angular momentum
(including the orbital part!) carried by the quarks \cite{ji:prl}
\bea
\fbox{\mbox{DVCS}}\quad \Leftrightarrow \quad
\fbox{\mbox{GPDs}}\quad \Leftrightarrow \quad {\vec J}_q.
\eea
However, even though GPDs clearly are interesting observables, it is not
{\it a priory} clear whether they have a simple physical interpretation
\cite{ji}.
In these notes, we will investigate
the physics of GPDs in the limit
where the momentum transfer on the target is purely transverse and we
will show that the $t$-dependence of GPDs in this limit can be identified
with the Fourier transform of parton distributions in the transverse plane
w.r.t. the impact parameter.

\section{Generalized parton distributions}
GPDs are defined very similar to
regular parton distributions
\bea
F_q(x,\xi,t)&=&\int \frac{dx^-}{4\pi} e^{ip^+x^- x}
\left\langle P^\prime
\left| \bar{q}\left(-\frac{x^-}{2}\right)\gamma^+ q\left(\frac{x^-}{2}\right)
\right| P\right\rangle \label{eq:def}\\
&=& H_q(x,\xi,t) \frac{1}{2p^+}\bar{u}(P^\prime)\gamma^+ u(P)
+ E_q(x,\xi,t) \frac{1}{2p^+}\bar{u}(P^\prime)
\frac{i\sigma^{+\nu}\Delta_\nu}{2M} u(P)
\nonumber \eea
where light front (LF) coordinates are defined as
$x^\pm=\frac{1}{\sqrt{2}}\left(x^0\pm x^3\right)$,
${\bf x}_\perp = (x^1,x^2)$, and $\Delta=p^\prime -p$.

Unlike the familiar (forward) parton distribution functions (PDFs)
\bea
q(x)&=&\int \frac{dx^-}{4\pi} e^{ip^+x^- x}
\left\langle P
\left|
\bar{q}\left(-\frac{x^-}{2}\right)\gamma^+ q\left(\frac{x^-}{2}\right)
\right| P\right\rangle \label{eq:defpdf},
\eea
which have the physical meaning of a momentum density in the
infinite momentum frame (IMF), GPDs do not have the meaning of
a density but that of a probability amplitude:, GPDs
describe the amplitude for finding a quark with momentum fraction
$x-\xi$ (in the IMF) in the nucleon and replacing it back into
the nucleon with a momentum transfer $\Delta^\mu$.
%, taking the overlap
%with a nucleon with momentum $p^\prime$.

Several limiting cases are familiar: for $p^\prime=p$ one recovers
the usual PDFs
\bea
H(x,0,0)=q(x)\quad \quad \quad \quad \quad
\tilde{H}(x,0,0)=\Delta q(x),
\eea
while integration over $x$ yields the familiar form factors, e.g.
\bea
\int dx H(x,\xi,t) = F_1(t) \quad \quad \quad\quad
\int dx E(x,\xi,t) = F_2(t) . \label{eq:form}
\eea
Therefore another simple physical interpretation is that
{\bf GPDs `measure' the contribution of
quarks with longitudinal momentum fraction $x$ to the corresponding
form factor}, i.e. GPDs are the form factor for the operator
that `filters out' quarks with longitudinal momentum fraction $x$
in the proton (in conventional PDFs, the quark is inserted back into
the nucleon without momentum transfer).

Given the latter analogy, and given the fact that one can associate the
Fourier transform of form factors with charge distributions in position space,
it is very tempting to expect that GPDs also contain some information
about the distribution of partons in position space.

\section{The physical interpretation of $H(x,0,t)$}
In this section we will focus on the helicity non-flip case (unpolarized)
\footnote{Note that our
results can be easily generalized to the polarized counterpart
$\tilde{H}(x,0,t)$.}. Furthermore,  we consider a situation when the
momentum transfer between the initial and final state in
Eq. (\ref{eq:def}) is purely transverse. In this limit,
$F_q(x,0,t)=H(x,0,t)$ can be written
as overlap integrals between LF wave functions
(Fock space amplitudes)\cite{diehl}
$\Psi_N(x,{\bf k}_\perp)$
\bea \!\!\!\!\!
H(x,-\!{\bf \Delta}_\perp^2)
= \sum_N \!\sum_j\!\!
\int\!\! \left[dx\right]_N \!\int\!\! \left[d^2{\bf k}_\perp\right]_N\!
\delta(x-x_j) \Psi^*_N(x_i,{\bf k}_{\perp,i}^\prime)
\Psi_N(x_i,{\bf k}_{\perp,i})
\label{eq:exact}
\eea
where
${\bf k}_{\perp, i}^\prime = {\bf k}_{\perp, i} - x_i {\bf \Delta}_\perp$ for
$i\neq j$ and
${\bf k}_{\perp, j}^\prime = {\bf k}_{\perp, j} +(1-x_j) {\bf \Delta}_\perp$.
Eq. (\ref{eq:exact}) is very
similar to the expression for the form factor in the Drell-Yan frame, except
that the $x$ of the `active' quark is not integrated over, and it is
exact if one knows the $\Psi_N$ for {\it all} Fock components.
The highly compact notation in Eq. (\ref{eq:exact}) is illustrated by
considering as an example the contributions from the two and three particle
Fock components to $H(x,{\bf q}_\perp)$ for the pion
\bea
H(x,-\!{\bf \Delta}_\perp^2) &&=
\int\!\! d^2{\bf k_\perp}
\psi^*_{\bf \Delta_\perp}(x,{\bf k_\perp}+{\bf \Delta_\perp})
\psi_{\bf 0_\perp}(x,{\bf k}_\perp) \label{eq:overlap}\\
+&&
\int \!\!
d^2{\bf k_{\perp}}d^2{\bf l_{\perp}}dy\,
\psi^*_{\bf \Delta_\perp}(x, {\bf k_{\perp}}+{\bf \Delta_\perp},
y,{\bf k_{\perp}})\psi_{\bf 0_\perp}
(x, {\bf k_{\perp}},y,{\bf l_{\perp}})
\,+\, .\, .\, .\, ,\nonumber
\eea
where `...' indicates contributions from 4 and more particle
Fock components.
\subsection{Nonrelativistic form factors and charge density in position
space}
Eq. (\ref{eq:overlap}) resembles expressions
for the  nonrelativistic (NR) form factor.
For example for NR two \& three body systems, the form factor reads
respectively
\bea
F({\vec q}) &=& \int d^3{\vec k}
\psi^*_{\vec q}({\vec k}+{\vec q})
\psi_{\vec 0}({\vec k})
\label{eq:ff1}\\
F({\vec q})&=&\int d^3{\vec k}_1d^3{\vec k}_2\,
\psi^*_{\vec q}({\vec k}_1+{\vec q},
{\vec k}_2)\,
\psi_{\vec 0}({\vec k}_1,{\vec k}_2).
\eea
Like GPDs,
$F({\vec q})$ is also off-diagonal in the momentum since the
form factor is the non-forward (${\vec P}^\prime \neq {\vec P}$)
matrix element of the vector current. In order to relate
$F({\vec q})$ to position space densities, one first uses that
NR boosts are purely kinematic, i.e.
\be
{\vec k}_i^\prime = {\vec k}_i+ x_i {\vec q},
\ee
with $x_i=\frac{m_i}{M}$, to express the boosted wave functions in
terms of wave functions in the rest frame
\bea
\psi_{\vec q}({\vec k})&=&\psi_{\vec 0}({\vec k}-
x{\vec q})\label{eq:boost}\\
\psi_{\vec q}({\vec k}_1,{\vec k}_2)&=&
\psi_{\vec 0}({\vec k}_1-x_1{\vec q},
{\vec k}_2-x_2{\vec q})\nonumber .
\eea
One can thus rewrite $F({\vec q})$ as an autocorrelation
of the ${\vec P}={\vec 0}$ wave function
\bea
F({\vec q})&=&\int d^3{\vec k}\,
\psi^*_{\vec 0}({\vec k}+(1-x){\vec q})\,
\psi_{\vec 0}({\vec k})\label{eq:ff2}\\
F({\vec q})&=&\int d^3{\vec k}_1d^3{\vec k}_2\,
\psi^*_{\vec 0}({\vec k}_1+(1-x_1){\vec q},
{\vec k}_2-x_2{\vec q})\,
\psi_{\vec 0}({\vec k}_1,{\vec k}_2),
\nonumber \eea
for two and three particle systems respectively.
Finally, upon Fourier transforming to position space, one can
turn the
autocorrelation function in Eq. (\ref{eq:ff2})
into a density. More precisely, one can
express $F({\vec q})$  in terms of the charge density
$\rho({\vec r})$ at distance ${\vec r}$
from the center of mass ${\vec R}_{CM}
\equiv \sum_i \frac{m_i}{M} {\vec r}_i$.
\be
F({\vec q}) = \int d^3r e^{-i{\vec q}\cdot {\vec r}}
\rho ({\vec r}).
\label{eq:ff3}
\ee

\subsection{\bf relevance for GPDs}
Purely transverse boosts in the LF frame also form a Galilean subgroup
\bea
x_i\longrightarrow  x_i \quad \quad \quad \quad
{\bf k}_{\perp, i}\longrightarrow  {\bf k}_{\perp,i} +
x_i{\bf q}_\perp ,
\label{eq:galilei}
\eea
where the momentum fractions $x_i$ play the role of the mass fractions
$\frac{m_i}{M}$ in the NR case. Therefore, LF Fock space amplitudes
transform under purely $\perp$
boosts very similar to the way NR wave functions transform. For example,
for the two and three body Fock components one finds
\bea
\psi_{\bf q_\perp}(x,{\bf k_\perp}) &=& \psi_{\bf 0_\perp}
(x,{\bf k}_\perp-x{\bf q}_\perp)\label{eq:LFboost}\\
\psi_{\bf q_\perp}(x,{\bf k_\perp},y,{\bf l}_\perp) &=&
\psi_{\bf 0_\perp}
(x,{\bf k}_\perp-x{\bf q}_\perp,y,{\bf l}_\perp-y{\bf q}_\perp)\nonumber .
\eea
Because of the very similar boost properties and the
very similar expressions for GPDs one the one hand (\ref{eq:overlap})
and NR form factors on the other hand (\ref{eq:ff1}), one can
proceed with GPDs in analogy with NR
form factors to express them in terms of a density.
First one expresses $H(x,{\bf q}_\perp)$ in terms of
autocorrelations of Fock space amplitudes in
the ${\bf p}_\perp={\bf 0}_\perp$ frame
\bea
H(x,-\!{\bf \Delta}_\perp^2) &&=
\int\!\! d^2{\bf k_\perp}
\psi^*_{\bf 0_\perp}(x,{\bf k_\perp}+(1-x){\bf \Delta_\perp})
\psi_{\bf 0_\perp}(x,{\bf k}_\perp) \label{eq:overlap99}\\
&&\!\!\!\!\!\!\!\!\!\!\!\!\!\!\!\!\!\!\!\!\!\!\!\!\!\!\!\!\!\!\!\!+
\int \!\!
d^2{\bf k_{\perp}}d^2{\bf l_{\perp}}dy
\psi^*_{\bf 0_\perp}(x, {\bf k_{\perp}}+(1-x){\bf \Delta_\perp},
y,{\bf k_{\perp}}-x{\bf \Delta_\perp})\psi_{\bf 0_\perp}
(x, {\bf k_{\perp}},y,{\bf l_{\perp}})
+ . . .,\nonumber
\eea
and then one
Fourier transforms the $\perp$ coordinates to
position space. In complete analogy with NR form factors this allows
one to express $H(x,t)$ as the Fourier transform of parton
distributions in the $\perp$ plane
\be
H(x,-{\bf \Delta}^2_\perp) = \int d^2 {\bf b}_\perp
e^{-i{\bf \Delta}_\perp\cdot {\bf b}_\perp} f(x,{\bf b}_\perp),
\ee
where
{\bf $f(x,{\bf b}_\perp)$ is the probability density to find
a quark with momentum fraction $x$ at $\perp$ distance
${\bf b}_\perp$ from the $\perp$ center of momentum
${\bf R}_\perp^{CM}\equiv \sum_i x_i {\bf r}_{i,\perp}$.}

Note that $f(x,{\bf b}_\perp)$ is gauge invariant: to prove this,
one uses that a formal operator definition of $f(x,{\bf b}_\perp)$,
which in LF gauge is given by \cite{mb}
\bea
f(x,{\bf b}_\perp)&=&\int \frac{dx^-}{4\pi} e^{ip^+x^- x}
\left\langle \psi_{loc}
\left|
\bar{q}\left(-\frac{x^-}{2},{\bf b}_\perp\right)\gamma^+
q\left(\frac{x^-}{2},{\bf b}_\perp\right)
\right| \psi_{loc}\right\rangle \label{eq:deff}.
\eea
Here $\left| \psi_{loc} \right\rangle \equiv
\int d^2 p_\perp \psi({\bf p}_\perp) \left| p \right\rangle$ is
a wave packet of plane wave proton states which is very localized
in the transverse direction, but still has a sharp longitudinal momentum.
This can be accomplished by taking $\psi({\bf p}_\perp)=const.$, which
guarantees that the $\perp$ center of momentum (the variable conjugate to the
total $\perp$ momentum!) is at the origin. To make Eq. (\ref{eq:deff})
manifestly gauge invariant one inserts a
straight line gauge string from $(-\frac{x^-}{2},{\bf b}_\perp)$
to$(\frac{x^-}{2},{\bf b}_\perp)$.

This example also illustrates why the $\perp$ center of momentum
is of significance here:
first of all, the intrinsic dynamics of the hadron
is independent of the overall $\perp$ momentum in the IMF ($\perp$
boosts are purely kinematical), which allows to localize the
the wave packet in the $\perp$ direction through a wave packet.
Since the variable conjugate to the overall $\perp$ momentum is
${\bf R}_\perp^{CM}$, localizing the wave packet is equivalent to
working in the frame where ${\bf R}_\perp^{CM}=0$.
\footnote{Of course the same reasoning also applies to the familiar
case of the NR form factor.}

\section{Discussion}
By means of a Fourier transform,
$H(x,-{\bf \Delta}^2_\perp)$ tells us how partons are distributed in
the transverse plane as a function of the distance from the
$\perp$ center of momentum. First of all this provides us, at least
for $\xi=0$, with a very simple interpretation for GPDs in terms
of a density.
A similar interpretation exists for
$\tilde{H}(x,-{\bf \Delta}^2_\perp)$. in terms of
polarized impact parameter dependent quark distributions in a longitudinally
polarized target.
These fundamental results by themselves are already important
observations because they illustrate the kind of physics that
one can learn about hadrons by studying GPDs.
However, above and beyond this physics insight, our
results also have a number of practical application because
they allow one to use geometric insights about hadron structure to
model GPDs as illustrated by the following examples:

\begin{itemize}
\item[1.] First of all, the mere fact that the ${\bf b}_\perp$ distribution
is measured w.r.t. ${\bf R}_\perp^{CM}$ necessarily implies that
the width of the ${\bf b}_\perp$ distribution goes to zero as
$x\rightarrow 1$, since the active quark becomes the $\perp$
center of mass in that limit. In momentum space this implies
that $H(x,t)$ should become $t$-independent as $x\rightarrow 1$.
Many models for GPDs are consistent with this kind of behavior, but
our analysis indicates that this is a general result.
\item[2.] Eq. (\ref{eq:deff}) illustrates that
the impact parameter dependent parton distributions
can be expressed in the form
\be
f(x,{\bf b}_\perp)\sim \left\langle \psi_{loc}\left|
b^\dagger(xp^+,{\bf b}_\perp)
b(xp^+,{\bf b}_\perp)\right|\psi_{loc}
\right\rangle
= \left|\left.b(xp^+,{\bf b}_\perp)|\psi_{loc}
\right\rangle\right|^2
,
\ee
where $b(xp^+,{\bf b}_\perp)$ creates quarks of
long. momentum $xp^+$ at $\perp$ position
${\bf b}_\perp$. This implies that
$f(x,{\bf b}_\perp)$
has a probabilistic interpretation,
and therefore it has to satisfy positivity
constraints, i.e.
\be
f_q(x,{\bf b_\perp}) \equiv \int d^2{\bf \Delta}_\perp
H_q(x,0,-
{\bf \Delta}_\perp^2)e^{i{\bf \Delta}_\perp\cdot {\bf b_\perp}}
\ee
should be positive for all $x$ and all ${\bf b_\perp}$ (for antiquarks the
same holds modulo an obvious sign).
This positivity condition provides us with a new constraint on all models
for GPDs.
\item[3.] Since very little is known about the actual $t$-dependence
of GPDs, it is often convenient to use simple parameterizations.
A commonly used form (motivated by LF constituent models) is
$H(x,0,-{\bf \Delta}_\perp^2) =
q(x) e^{-a {\bf \Delta}_\perp^2 \frac{1-x}{x}}$.
However, such a ${\bf \Delta}_\perp$-dependence
yields an unacceptably rapid
growth for the $\perp$ hadron size $\langle {\bf b}_\perp^2 \rangle
\sim \frac{1}{x}$ in the small $x$ region.
Space time descriptions of parton structure in the small
$x$ region\cite{gribov} suggest a $\perp$ size of hadrons that
grows like $\alpha \ln \frac{1}{x}$. Translated into GPDs this implies that
the small $x$ behavior should be parameterized by a functional form like
\be
H(x,0,-{\bf \Delta}_\perp^2) = q(x)
e^{-\alpha {\bf \Delta}_\perp^2 \ln \frac{1}{x}}
\quad  \mbox{or} \quad
%H(x,0,-{\bf \Delta}_\perp^2) =
q(x)
e^{-\alpha {\bf \Delta}_\perp^2 (1-x)\ln \frac{1}{x}}
\ee
(the $2^{nd}$ ansatz is also consistent with the
Drell-Yan-West relation).
\item[4.] At a more qualitative level, one expects quarks at large $x$
to come from the more localized valence `core' of the hadron, while
the small $x$ region should also receive contributions from
the much wider meson `cloud'
and therefore one would in general expect a gradual increase of the
$t$-dependence of $H(x,0,t)$ as one goes from larger to smaller values of
$x$
\end{itemize}
\section{The Physics of $E(x,0,t)$}
So far we have only considered GPDs in an
unpolarized (or longitudinally) polarized nucleon. For polarized
nucleon states (we use an IMF helicity basis here \cite{sjb}) one finds
the following useful relations when $\Delta^+=0$
\bea
\int \frac{dx^-}{4\pi} e^{ip^+x^- x}
\left\langle P+\Delta, \uparrow
\left| \bar{q}\left(0\right)\gamma^+
q\left({x^-}\right)
\right| P,\uparrow\right\rangle
&=& H(x,0,-{\bf \Delta}_\perp^2) \label{eq:flip}\\
\int \frac{dx^-}{4\pi} e^{ip^+x^- x}
\left\langle P+\Delta, \uparrow
\left| \bar{q}\left(0\right)\gamma^+
q\left({x^-}\right)
\right| P,\downarrow\right\rangle
&=& -\frac{\Delta_x-i\Delta_y}{2M}E(x,0,-{\bf \Delta}_\perp^2)
\nonumber ,
\eea
i.e. if we take for example a nucleon polarized
in the $x$ direction (in the IMF)
\be
F_q(x,0,-{\bf \Delta}_\perp) = H(x,0,-{\bf \Delta}_\perp^2) + i
\frac{\Delta_y}{M} E(x,0,-{\bf \Delta}_\perp^2)
\label{eq:fsx} .
\ee
%and similar for other polarization directions.
Eq. (\ref{eq:fsx}) allows us to
draw the following conclusions for the physics of $E(x,0,t)$
\begin{itemize}
\item[1.] We know already that
$H(x,0,-{\bf \Delta}_\perp^2)$ describes the distribution of
unpolarized partons in the $\perp$ plane for a nucleon that is
either unpolarized or polarized in the $z$ direction. From Eq.
(\ref{eq:fsx}) we can immediately read off that
{\bf $\frac{\Delta_\perp}{M} E(x,0,-{\bf \Delta}_\perp^2)$
describes how the distribution of partons in the $\perp$ plane
depends on the polarization of the nucleon}. For example, for
a nucleon polarized in the $x$ direction
the z-momentum distribution of partons on the +y side will
differ from the one on the $-y$ side
--- which makes sense if one considers the classical picture of a
sphere that spins around the x-direction while moving in the z-direction.
The z-momenta due to spinning add to the
translatory momenta for `partons' on the $+y$ side and subtract on
the $-y$ side and hence it is not surprising to find a
difference between the parton distributions on the $\pm y$ sides.
\item[2.] Since the probabilistic interpretation should not depend on the
polarization, above interpretation provides us with another
positivity constraint relating the Fourier transform of
$E$ to the one of $H$:
\be
\!\!\!\!\!\!\!\!\!\!\!\!
\left| \frac{{\bf \nabla}_{b_\perp}}{M}
\int d^2{\bf b}_\perp e^{i{\bf b}_\perp \cdot {\bf \Delta_\perp} }
E(x,0,-{\bf \Delta}_\perp^2) \right|
<
\int d^2{\bf b}_\perp e^{i{\bf b}_\perp \cdot {\bf \Delta_\perp} }
H(x,0,-{\bf \Delta}_\perp^2) .
\ee
\item[3.] Finally, we are also able to illustrate the physics of Ji's angular
momentum sum rule $\langle J_q \rangle = \frac{1}{2}\int dx x\left[
H_q(x,0,0)+E_q(x,0,0) \right]$. For this purpose, we first note
that GPDs for $\xi=0$ allow the simultaneous determination of
the momentum of partons in the $z$ direction and
their position in the $\perp$ direction
(which is very much reminiscent of angular momentum
$L_x=xp_z-zp_x$) and it should thus not surprise that GPDs
allow determining the total angular momentum of the quarks.
In order to illustrate this connection, let us consider a
nucleon polarized in the $x$ direction.\footnote{I was only able to
find a simple physical interpretation for the connection between
GPDs and the $\perp$ components of ${\vec J}_q$. However, by
rotational invariance, one can apply the result also to $J^z_q$, and,
by boost invariance, $J_q^z$ should also be he same for a nucleon
moving in the $z$-direction.}
Because we are interested in the
angular momentum in the rest frame of the target, we again
need to consider again nucleons polarized in the $x$-direction, but now
in the nucleon rest frame. This differs from the IMF by a Melosh
rotation and one finds up to terms linear in $\Delta_\perp$
\be
\!\!\!\!\!\!\!\!\!\!\!\!
F_q(x,0,{\bf \Delta}_\perp) = H(x,0,-{\bf \Delta}_\perp^2) + i
\frac{\Delta_y}{2M} \left[  H(x,0,-{\bf \Delta}_\perp^2)+
E(x,0,-{\bf \Delta}_\perp^2)\right]
\label{eq:fsxrf} .
\ee
For the $x$ component of the angular momentum carried by the quarks
one thus finds
\footnote{In order to calculate $\langle J_x\rangle$, it first seems that one needs both
$\langle yT^{0z}\rangle$ and $\langle zT^{0y}\rangle$, but for
stationary localized states the two expectation values are (up to a
sign) the same and
it is thus sufficient to calculate $\langle yT^{0z}\rangle$.
Furthermore, in the rest frame one can replace the usual momentum
density in  $\langle yT^{0z}\rangle$ by the LF momentum density
$T^{++}$, whose matrix elements can be expressed by the GPDs.}
\bea
\!\!\!\!\!\!\!\!\!\!\!\!\!\langle J^x_q\rangle &=& \langle yT^{0z}\rangle - \langle
zT^{0y}\rangle =  2\langle yT^{0z}\rangle
= 2\frac{M}{2}\int dx \int d^2{\bf b}_\perp \, b_y xf(x,{\bf
  b}_\perp) \\
&=& M i\left.\frac{d}{d\Delta_y}
\int dx\, x
F_q(x,0,{\bf \Delta}_\perp) \right|_{\Delta=0}
\nonumber
=\frac{1}{2}\int dx\, x\left[H(x,0,0)+E(x,0,0)\right]
\eea
which is the angular momentum relation derived in Ref.
\cite{ji:prl}. What is new is that through the
geometric interpretation of GPDs, we are now able
to understand the physics of this sum rule
by relating GPDs to the expectation value of
$\langle yp_z\rangle$ for a nucleon polarized
in the $x$ direction.
\end{itemize}

\section*{Acknowledgments}
This work was supported by a grant from DOE (FG03-95ER40965) and through
Jefferson Lab by contract DE-AC05-84ER40150 under which the Southeastern
Universities Research Association (SURA) operates the Thomas Jefferson
National Accelerator Facility.

\end{document}